# Spectroscopy of atmospheric pressure air jet plasma in transverse arc discharge

Valeriy Chernyak[1], Vadym Naumov[2], and Vitaliy Yukhimenko[1]

[1] Radiophysics Faculty, Taras Shevchenko Kiev National University,
Prospect Acad. Glushkova 2/5, Kiev 03122 Ukraine, e-mail: chern@univ.kiev.ua

[2] Institute of Fundamental Problems for High Technology, Ukrainian Academy of Sciences,
Prospect Nauki 45, Kiev 03028 Ukraine, e-mail: naumov@ifpht.kiev.ua

## Abstract

Spectroscopic characterization of a specific case of the atmospheric pressure air jet plasma in the transverse cw dc arc discharge of high voltage was done. Within the spectrum of wavelengths from 200 to 1100 nm all remarkable emissions were monitored, and all excited atomic lines of N, O, H and molecular bands of NO, $N_2$, $O_2$, OH, CO, CN were identified. Using relative intensities of analytical CuI lines 510.5, 515.3, 521.8 nm (the product of electrode emission) and $N_2$ ($C^3\Pi_u$-$B^3\Pi_g$) $2^+$-system band at 337.1 nm (the dominating component of plasma-forming gas), the temperature of electronic excitation of free atoms, $T_{exc}$, and the temperatures of excitation of vibrational and rotational states of molecules, $T_V$ and $T_R$, were determined. It was found that there is no local LTE in this arc discharge air plasma during its space/time evolution, and effects of strong non-izothermality have a place in this case.

***Keywords:*** Air plasma; Arc discharge; Nonequilibrium, Non-izothermality, Spectroscopy

## Introduction

Spectroscopic characterization of air plasma in electric discharges is of permanent interest in many labs because of various important applications [1]. From the optical spectra emitted by the air plasma one can deduced all basic plasma parameters and its state [2]: composition of chemical elements (according to emitted lines), concentrations of excited atoms and molecules upon energetic levels (from the intensity of spectral lines), densities of charged and neutral particles (from the broadening of spectral lines), temperatures of excitation of atomic and molecular states (from the relative or absolute spectral line intensities). The main condition is a correct use of the spectroscopic diagnostics in every specific case, e.g. application of approximation of the optically thin plasma in conditions of high atmospheric pressure and approximation of the local thermodynamic equilibrium (LTE) in conditions of high non-izothermality when the characteristic temperatures of different plasma components may be differenced within the relation: electron temperature $T_e$ > vibrational temperature $T_V$ > rotational temperature $T_R \geq$ translational gas temperature $T_g$ [3].

Among possible types of the nonequilibrium high-pressure discharges: spark, corona, barrier, etc [4], one specific case is very interesting. This is a transverse arc in a blowing flow with a stationary current column or with a rotating one in a vortex flow. It is an intermediate case of the high-voltage low-current self-sustained discharge with a self-adjustable arc supported by the plasma flow [5]. It differs from the non-stationary gliding arc of Czernichowski type [6, 7] by the fixed arc length. It also has a convective cooling of the plasma column by the air flow but without conductive heat losses at the walls since it is a free arc jet. An intensive transverse ventilation of the arc plasma increases its ionization non-equilibrium and non-izothermality [8]. This type of the arc was tested successfully in different variants with the primary and secondary discharges for the plasma-assisted processing of various homo- and heterophase gas and liquid substances [9, 10]. However, despite of achievements in practical applications, there are still enough issues for research. One of the main points is a mechanism of the transition from the quasi-equilibrium to non-equilibrium arc, i.e. from the thermal to non-thermal ionization. In this paper, we present results of our spectroscopic characterization of the air plasma in such arc discharge in order to get more deep understanding in physics of nonequilibrium processes.

## Methodology

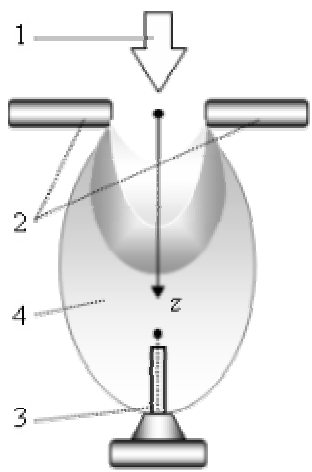

**Fig.1**. Scheme of transverse blowing arc discharge: 1 is air nozzle, 2 are electrodes, 3 is probe, 4 is plasma jet.

Experiments were done for a scheme of the transverse blowing arc as shown in Fig.1. A free jet of the atmospheric air ran from the nozzle across two horizontal opposite electrodes and formed a bright crescent-shaped electric arc as well as a highly reactive afterglow. We used the rod electrodes with diameter $d = 5$ mm. A nominal gap between the electrodes from which we started usually was $\delta = 1$ mm. Since the electrodes were not cooled, the electric discharge energy was transferred totally to the air plasma flow. We applied electrodes made from different materials: copper and graphite, in order to see a spectroscopic difference. The air nozzle was axisymmetric, with the inner diameter $\varnothing = 1$ mm, made from the stainless steel. It was maintained vertically perpendicular to the electrode axis at the length $L = 5\text{-}10$ mm and it was centered strictly between the electrodes. We used a standard technical dry air system supply with the flow meters. It was the high enough gasdynamic pressure in the flow to blow out the electric arc downstream. In fact, we can regulate the arc discharge geometry as by the gap $\delta$ between the electrodes and by the length $L$ between the nozzle exit and the electrodes. The last allows to control both the air blowing of the arc and the air cooling of the electrodes. Then, we can regulate the air flow rate, $G$, and arc discharge current, $I_d$. The arc discharge is powered by the DC source at the ballast resistance $R = 2$ kΩ in the circuit. Current-voltage parameters were measured with the standard electronics. For optical diagnostics, a classic method of the emission UV-NIR spectroscopy was applied. The plasma radiation was measured by two means: 1) portable rapid PC-operated CCD-based multi-channel optical spectra analyzer (MOSA), which has a wide wavelength survey within 200-1100 nm but medium spectral resolution ~0.2 nm, and 2) spectral combine KSVU-23, including the scanning monochromator DMR-2, PMT detector FEU-100 and PC recorder, which provides a high spectral resolution up to 0.01 nm but slow scanning speed. Measurements were conducted in different cross-sections along the arc and afterglow. The spatial resolution was about 0.1 mm. The images were normally focused by the quartz lens at the bench 5-focus distance from the arc directly on the entrance slit of the spectral device. In case of the MOSA, we used a fine optical fiber with a microlens. For calibration, a set of etalon spectral sources, including the mercury, deuterium, xenon and tungsten lamps were applied.

Within the available spectrum of wavelengths from 200 to 1100 nm we conducted a monitoring of all remarkable emissions and we identified all basic atomic lines of N, O, H as well as molecular bands of NO, $N_2$, $O_2$, OH, CO, CN, which we were able to recognize [11, 12]. For qualitative diagnostics, we utilized relative intensities of CuI lines (product of electrodes' emission) at 510.5, 515.3, and 521.8 nm in order to determine the temperature of electronic excitation of atoms, $T_{exc}$, and we used $N_2$ $C^3\Pi_u$-$B^3\Pi_g$ $2^+$-system bands (as a dominating component in air plasma) at 337.1 nm (0-0), 357.7 (0-1), 380.5 (0-2), etc in order to determine the temperature of excitation of vibrational states of molecules, $T_V$, by the Ornstein method as a commonly accepted approach [13]. It is assumed the plasma is optically thin and the spectral lines are not self-absorbed. The temperature of excitation of rotational states, $T_R$, because of non-resolved rotational spectral structure, was estimated by the comparison of the measured spectra of $N_2$ $2^+$(0,0) band at 337.1 nm and the corresponding synthetic spectra calculated on the known spectral constants for $N_2$ $C^3\Pi_u$-$B^3\Pi_g$ transition, using the Gauss-like instrumental function as usual. To increase the accuracy of measurements, a statistical processing was performed. The error in the emissivity measurements did not exceed 15%, and the temperatures were determined with an uncertainty ±10%. On this base, we draw the curves of changes of the specific emission intensities and characteristic temperatures along the plasma, depending on the arc discharge power and the airflow rate. The overall studies were conducted within the range of variations $I_d$ from 100 to 500 mA and $G$ from 40 to 180 cm$^3$/s (airflow velocities 40-180 m/s).

## Results and Discussions

A transverse arc discharge in the airflow was ignited with a high voltage at the shortest distance between the electrodes that corresponds to breakdown when the electric field reached ~3 kV/mm [4]. Under the action of the gasdynamic pressure of the incident airflow, the electric arc was forced to bow down and elongated along the stream, so that the current increased and the voltage dropped down a little bit. The air flow led not only to bending and to blowing of the arc current channel but also to stabilization of the plasma column due to the convective withdrawal of the energy (radiative losses are neglected). Due to a high-speed flow, the air plasma had to be turbulized, and it additionally contributed to the suppression of ionization-overheating instability. Thus, gasdynamics and convective heat/mass transfer favored the steady-state arc burning. Fig.2 shows a typical spectrum of the UV-VIS-NIR emission of the air plasma in the studied arc discharge as registered by the MOSA. It is rich of spectroscopic information. We recognized here the nitride oxide NO γ-system ($A^2\Sigma^+$-$X^2\Pi$: (0-0) 226.9 nm, (0-1) 236.3 nm, (0-2) 247.1 nm, etc); hydroxyl OH UV system ($A^2\Sigma$-$X^2\Pi$: (0-0) 306.4-308.9 nm); oxygen $O_2$ Shumann-Runge bands ($B^3\Sigma^-_u$-$X^3\Sigma^-_g$: (0-14) 337.0 nm); nitrogen $N_2^+$ 1$^-$ system ($B^2\Sigma^+_u$-$X^2\Sigma^+_g$: (1-0) 358.2, (1-1) 388.4, (0-0) 391.4 nm, etc); $N_2$ 2$^+$ system ($C^3\Pi_u$-$B^3\Pi_g$: (0-0) 337.1, (0-1) 357.7, (0-2) 380.5, (1-0) 316.0 nm, etc); and even week $N_2$ 1$^+$ system ($B^3\Pi_g$-$A^3\Sigma^-_u$: 570-750 nm). Among atomic lines, we recognized HI Ballmer α line 656.3 nm, OI lines (777.3, 844.6, 926.0 nm), and NI lines (746.8, 818.8, 868.3 nm). There are a lot of Cu lines due to evaporation of copper electrodes, but the most strong CuI lines 324.7 and 327.4 nm were overlap with the $N_2^+$ 1$^{(-)}$ bands, therefore we used the CuI lines 510.5, 515.3, 521.8, and 578.2 nm. The interference of the $N_2$ 2$^+$ system also precluded the diagnostics of the OH (A-X) band at 308 nm. Thus, the most results below were obtained with the $N_2$ 2$^+$ bands only.

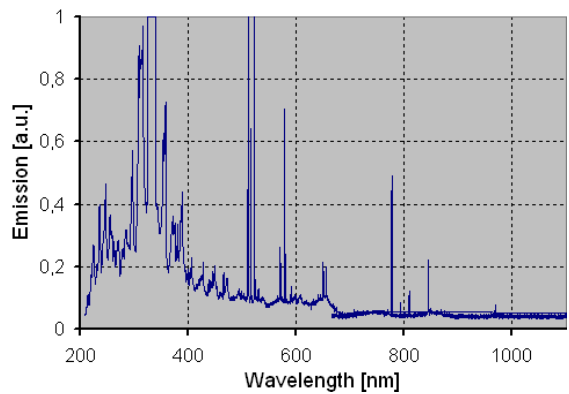

**Fig.2**. Emission of air plasma flow in the DC transverse blowing arc discharge. $U_d$ = 1.2 kV, $I_d$ = 200 mA, $P$ = 1 atm, z = 7 mm.

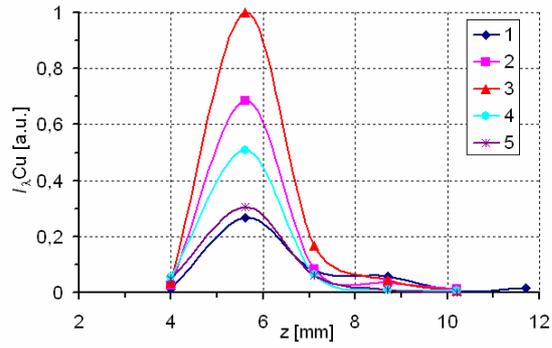

Fig.3

A distinguishing feature of the plasma column in the transverse blowing arc discharge is their curvature along the flow. Therefore, all dependencies of the emission intensities and spectral distributions along the z-axe downstream are of non-linear character. Fig.3 shows typical dependencies of the emissivity $I_\lambda(z)$ for the CuI line 510.5 nm. Here, curves 1-5 correspond to air flow rates $G$ = 40, 80, 110, 150, 180 m/s at the discharge current $I_d$ = 200 mA. As is seen, the absolute emissivity $I_{\lambda Cu}$ changes very much with the flow rates $G$ but the position of $I_\lambda^{max}$ at the distance z = 5.5 mm and the FWHM value ≈ 2 mm for all curves are almost constant. It tells about the stability of the arc current channel in the range of the used flow rates at the given arc discharge geometry. The effect of the "ignition" of the luminescence of the excited atoms downstream the arc tells about the overheating of electrons resulted from the kinetic non-equilibrium conditions in the gasdynamically moving and convectively cooling air plasma. As is seen, the absolute emissivity $I_{\lambda Cu}$ changes very much with flow rates $G$ but the position of $I_\lambda^{max}$ at z ≈ 5.5 mm and FWHM value ≈ 2 mm for all curves are almost constant. It tells about the stability of the arc current channel in the range of used flow rates at the given arc discharge geometry. The effect of "ignition" of the luminescence of excited Cu atoms downstream the arc tells about the overheating of electrons resulted from the kinetic non-equilibrium conditions in the gasdynamically moving and convectively cooling air plasma.

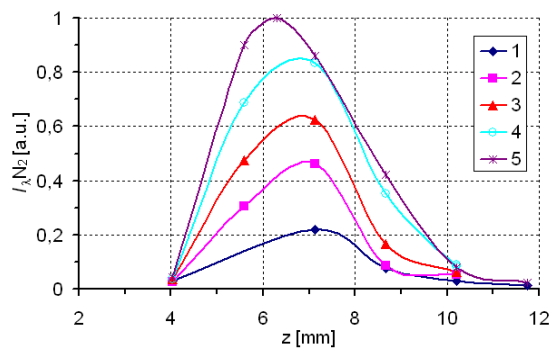

Fig.4

The distribution of the emission of the excited $N_2$ molecules along the z-axe has also a non-monotonic character. Fig.4 shows typical dependencies of the emissivity $I_\lambda(z)$ for the $N_2$ $2^+(0-0)$ band 337.1 nm. Here, curves 1-5 also correspond to air flow rates $G$ = 40, 80, 110, 150, 180 m/s at the discharge current $I_d$ = 200 mA. The comparison of $I_{\lambda Cu}(z)$ and $I_{\lambda N2}(z)$ in Fig.3 and Fig.4 tells that $I_{\lambda N2}(z)$ distributions are sufficiently larger and are shifting downstream relatively to $I_{\lambda Cu}(z)$ curves. With the above-mentioned conclusions about the arc current channel and plasma dynamics, it evidences that the air plasma afterglow is conditioned by the nonequilibrium emission of the excited $N_2$ molecules.

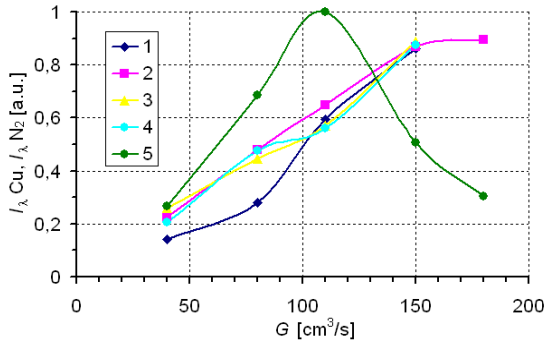

**Fig.5**

With the discharge current $I_d$ the absolute emissivity $I_{\lambda N2}$ (337.1 nm) increased (by few tens percent) but it reacted more strongly (in direct proportion) to increase in the flow rate $G$. It illustrates in Fig.5 by the curves 1-4, taken for the different values $I_d$ from 100 to 400 mA at $z \approx 5.5$ mm (see also Fig.4). In comparison with the almost liner dependence $I_{\lambda N2}(G)$, the dependence $I_{\lambda Cu}(G)$ is non-linear. This is clear shown by the curve 5 in Fig.5, given for the emissivity $I_{\lambda Cu}$ (510.5 nm) at the same values $I_d = 200$ mA and $z \approx 5.5$ mm (see also Fig.3). The $I_{\lambda Cu}$ functional extreme is at $G = 110$ m/s. Such behavior may be explained by the those that, on the one side, the increasing $G$ at the fixed $I_d$ should lead to increasing of the level of nonequilibrium of the air plasma due to acceleration of the energy carry-over from the arc current channel, and therefore to increasing of the emissivity of the excited Cu atoms. On the other side, with increasing of the airflow the erosion of electrodes and therefore the concentration of Cu atoms in the discharge decreased.

On the base of the measured values $I_{\lambda Cu}$ and $I_{\lambda N2}$, using the Boltzmann plot, we determined the corresponding temperatures of electronic excitation of Cu atoms, $T_{exc}$, and vibrational excitation of $N_2$ molecules, $T_V$. Due to a relatively high degree of ionization ($n_e/n > 10^{-3}$) and atmospheric pressure, we considered further that $T_{exc} \approx T_e$ as usual for such plasma [14]. As expected, $T_e$ and $T_V$ differenced very much, so that $T_e/T_V >> 1$ everywhere. At that, the level of non-izothermality is not permanent along the plasma flow. It depends not only on the current $I_d$ of the arc discharge but also on the airflow rate $G$ that is blowing the arc plasma column, providing convective heat/mass transfer. Especially large differences occur in the afterglow zone. Fig.6 shows typical dependencies $T_e(z)$ and $T_V(z)$ along the discharge at $I_d = 200$ mA for the different $G = 40-180$ m/s. As is seen, along the flow the value $T_e$ is 0.7-0.6 eV while $T_V$ is 0.4-0.35 eV. The highest value $T_e \approx 1.5$ eV is in the center of the arc. In the afterglow, $T_e$ decreases while $T_V$ keeps longer. Then increasing the current $I_d$, the temperature $T_e$ becomes larger. At a lager flow rate $G$, the gradient $T_e$ becomes smaller.

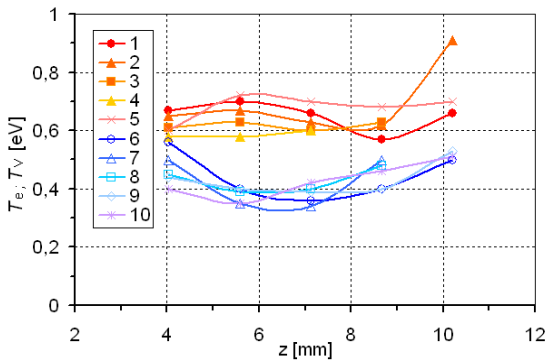

**Fig.6**

The non-equilibrium of the air plasma in the blowing arc discharge follows also from the estimation of the rotational temperature $T_R$ obtained at the same conditions. Fig. 7 shows the results of simulation of the V-R spectra for the $N_2$ $2^+$(0-0) band 337.1 nm, calculated at different $T_R = 0.05 \div 0.5$ eV with the step of 0.05 eV (from the curve 1 that is $T_R = 0.05$ eV to the curve 10 that is $T_R = 0.5$ eV) as compared with the measured data at $I_d = 200$ mA for $G = 40$ m/s (curve 11) and 80 m/s (curve 12) at the distance z $\approx$ 7 mm. Our estimation of $T_R$ is 0.2–0.25 eV. It differences from $T_V$ and $T_e$ more then twice. This evidences about real strong non-isothermality in the afterglow.

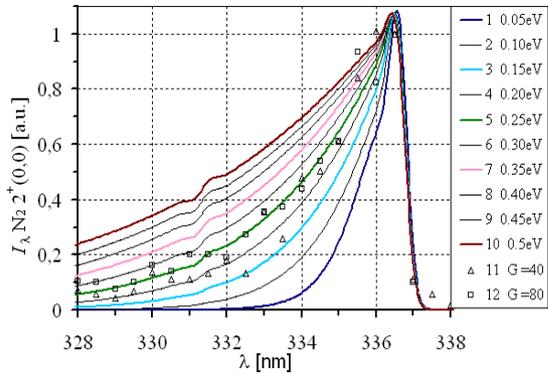

**Fig.7**

Thus, during the evolution of the air plasma in the transverse arc discharge the characteristic temperatures varied very much. However, with the exception of the afterglow, within the arc zone the gas and electron temperatures are coupled at the level of $T_g \sim$ 2000-3000 K and $T_e \sim$ 8000-12000 K, respectively. This indicates that we have the same transitional regime of the blowing arc discharge as in the gliding arc [7] when it is supported both by the thermal ionization (function of the gas temperature) and by the direct electron impact ionization (function of the electric field). It is particularly remarkable that despite of high pressure, the air plasma in the blowing arc discharge remains ionizationally non-equilibrium with the overheated electron component due to the effective convective heat carry-over.

**Conclusion**

We see that a high-voltage low-current arc discharge in the transverse air flow of atmospheric pressure can be a source of strong non-isothermal plasma with a high level of ionization. We found that there is no local LTE in this arc discharge air plasma during its space/time evolution, and the measured/estimated characteristic temperatures are within a relation $T_e \sim T_{exc} > T_V > T_R \sim T_g$, where the temperature of electronic excitation of free atoms $T_{exc}$ differences from the temperature of vibrationally and rotationally excited molecular states more then twice. Therefore, a conventional two-temperature approach with $T_e$ for electrons and $T_g$ for gas components is not valid in this case. Another character effect is an "ignition" of the plasma luminescence downstream the arc discharge resulted from the kinetic non-equilibrium conditions. The highest value $T_e \sim 1.5$ eV is measured in the center of the arc. In the afterglow zone, $T_e$ decreases rapidly while $T_V$ and $T_R$ keeps longer. The factors, which effects on the plasma nonequilibrium, are not only electric parameters of the arc discharge but also the gas dynamics and convective heat/mass transfer in the plasma flow.

In general, the results of our studies allow to conclude that even small variations in plasma conditions due to their spatial and temporal instability, decomposition, contamination, etc can produce large visual changes in the spectral emission which is functionally related with the temperatures and concentrations of plasma components, so the molecular spectroscopy could serve as a powerful tool for characterization of the nonequilibrium air plasma.

## Acknowledgments

The work is partially supported by the Ukrainian Ministry of Science and Education and by the Ukrainian State Foundation for Basic Research.